\documentclass[12pt,hidelinks]{article} 
\usepackage[sectionbib,sort]{natbib}
\usepackage{array,epsfig,fancyheadings,rotating}

\usepackage{sectsty, secdot}
\sectionfont{\fontsize{12}{14pt plus.8pt minus .6pt}\selectfont}
\renewcommand{\theequation}{\thesection\arabic{equation}}
\subsectionfont{\fontsize{12}{14pt plus.8pt minus .6pt}\selectfont}

\usepackage{amsmath}
\usepackage{amssymb}
\usepackage{amsfonts}
\usepackage{multirow}
\usepackage{amsthm}

\usepackage{bm, enumerate,xcolor}
\usepackage{graphicx,multicol}
\usepackage{float,subcaption}
\usepackage{multirow}
\usepackage{epstopdf}
\usepackage[linesnumbered,boxed]{algorithm2e}

\graphicspath{{sim_plot/}}

\usepackage{xr}
\allowdisplaybreaks

\newtheorem{example}{Example}
\newtheorem{lemma}{Lemma}

\theoremstyle{definition}

\begin{document}


\renewcommand{\baselinestretch}{2}

\markright{ \hbox{\footnotesize\rm Statistica Sinica
}\hfill\\[-13pt]
\hbox{\footnotesize\rm
}\hfill }

\markboth{\hfill{\footnotesize\rm FIRSTNAME1 LASTNAME1 AND FIRSTNAME2 LASTNAME2} \hfill}
{\hfill {\footnotesize\rm FILL IN A SHORT RUNNING TITLE} \hfill}

\renewcommand{\thefootnote}{}
$\ $\par


\fontsize{12}{14pt plus.8pt minus .6pt}\selectfont \vspace{0.8pc}
\centerline{\large\bf Binary Outcome Copula Regression Model}
\vspace{2pt} \centerline{\large\bf with Sampling Gradient Fitting}
\centerline{Weijian Luo$^1$, Mai Wo$^{2}$} 
\centerline{\it School of Mathematical Sciences$^1$, }\-\\[-10mm]
\centerline{\it National School of Development,
	$^2$}
\centerline{\it Peking University}
\vspace{.55cm} \fontsize{9}{11.5pt plus.8pt minus
.6pt}\selectfont


\begin{quotation}
\noindent {\it Abstract:Use copula to model dependency of variable extends multivariate gaussian assumption. In this paper we first empirically studied copula regression model with continous response. Both simulation study and real data study are given. Secondly we give a novel copula regression model with binary outcome, and we propose a score gradient estimation algorithms to fit the model. Both simulation study and real data study are given for our model and fitting algorithm.}

\vspace{9pt}
\par
\end{quotation}\par

\def\thefigure{\arabic{figure}}
\def\thetable{\arabic{table}}

\renewcommand{\theequation}{\thesection.\arabic{equation}}

\fontsize{12}{14pt plus.8pt minus .6pt}\selectfont

\setcounter{section}{0} 
\setcounter{equation}{0} 

\lhead[\footnotesize\thepage\fancyplain{}\leftmark]{}\rhead[]{\fancyplain{}\rightmark\hspace{1cm} \footnotesize\thepage}

\section{Introduction}\label{sec:intro}

Copula has been a powerful mathematical tool for modelling dependence structure of variables in last decades. Assume $\textbf{X}=(X_1,...,X_d)^T$ be a random vector of dimension $d >= 1$ and $Y$ be a random variable of our most concern, which means we take $Y$ as our response. Further assume $X_i$ each has culumlative distribution $F_i$ and density $f_i$ while $Y$ has cumulative distribution $F_0$ and density $f_0$. A copula function is defined as $C_\theta(u_1,...,u_d,v) = \mathbf{P}(F_1(X_1)\leq u_1,...,F_d(X_d)\leq u_d, F_0(Y)\leq u_0)$. It is quite clear that copula function is joint cumulative function with uniform marginal distribution. Assume the copula function of $(X_1,...,X_d,Y)$ is $C_\theta(u_1,...,u_d,v)$.Sklar's theorem claim joint cumulative function of $(X_1,...,X_d,Y)$ can be expressed via a composition of a copula function and marginal cumulative function, which means $$\mathbb{P}(X_1\leq x_1,...,X_d\leq x_d,Y\leq y) = C_\theta(F_1(x_1),...,F_d(x_d),F_0(y))$$. Copula formulation give a clear seperation of marginal distribution and dependence structure. Use of Copula has been found in wide variaty of applied science such like quantitative risk control \cite{Wu2011} and statistical modelling. Intense research on copula based statistical method has been proposed. \cite{Frechet1951} introduced copula at the first time. \cite{Sklar1959} give a theoretical background for copula. Most classical results on copula can be found in \cite{Nelsen1999} . \cite{Galiani2003} propose a copula application for financial derivative products management. \cite{Parsa2011} and \cite{Noh2013} propose a regression model based on copula, which they give name as copula regression.Even in recent years, copula based regression method have been continously proposed. \cite{Rainer2012} and \cite{Radice2016} have studied copula regression for binary outcomes.

Joint distributions with special copulas have shown properties different from joint gaussian assumption. So copula has brought researcher a good view to go beyond joint gaussian distribution. Interesting properties like tail dependency of copula has motivated statistical community to persistently do research on copula. Since copula has been used sucessfully in quantitative risk control. Most work focus on simple modelling dependency and extreme behavior of multivariate variables using copula, few have tried to use copula to do regression inference or classification prediction.A interesting idea is to use copula based method to do regression inference. \cite{Parsa2011} proposes a copula regression method. \cite{Noh2013} propose a copula-based regression method and analyze the asymptotic property. In this paper, will first give a implemention of copula based regression model, and thus we will give and analyze a copula based model estimation with binary outcomes based on latent variable model corresponding real data experiment.

\section{Preliminaries}\label{sec:model}

\subsection{Copulas and Backgrounds}\label{sec:glm}
Assume we have variables $(X_1,...,X_d,Y)$. Assume variables$\{X_i\,1\leq i\leq d\}$ has dependence with response variable $Y$ while each covariate $X_i$ and $X_j$ also has dependence to each other. A elegent formulation of the model is to give a copula dependence among all variables. Assume $(X_1,...,X_d,Y)$ has Copula $C_\theta(u_1,...,u_d,v)$ which represents variables dependence. Assume each variable has marginal distribution$F_1,...,F_d,F_0$. Joint cumulative distribution function is naturally 
$$\mathbb{P}(X_1\leq x_1,...,X_d\leq X_d,Y\leq Y) = C_\theta(F_1(x_1),...,F_d(x_d),F_0(y))$$
Before we go further, we give some basic lemma to demonstrate the model.
\begin{lemma}
Assume $(X_1,...,X_d,Y)$ has copula smooth $C_\theta(u_1,...,u_d,v)$, and marginal cumulative distribution(density) $F_i(f_i)$, then $(X_1,...,X_d,Y)$ has joint density:
$$f(x_1,...,x_d,y) = c_\theta(F_1(x_1),...,F_d(x_d),F_0(y))\Pi_{i=1}^d f_i(x_i)$$
where $c_\theta(u_1,...,u_d,v) = \frac{\partial^{d+1}C_\theta(u_1,...,u_d,v)}{\partial u_1\partial u_2 ... \partial v}$
\end{lemma}
The lemma gives a relation from cumulative distribution function and density function;
\begin{lemma}
Assume $(X_1,...,X_d,Y)$ has copula smooth $C_\theta(u_1,...,u_d,v)$, and marginal cumulative distribution(density) $F_i(f_i)$, then conditional mean of variable $Y$ given covariates $\vec{X} = \vec{x}$ is:
$$m(x) = \mathbb{E}(Y|X) = \frac{\int yc_\theta(F_1(x_1),...,F_0(y))\Pi f_i(x_i)f_0(y)dy}{C_X(F(\vec{X}))}$$
where $C_X(u) = \frac{\partial^{d}C_\theta(u_1,...,u_d,v=1)}{\partial u_1\partial u_2 ... \partial u_d}$
\end{lemma}

For many copula families, $m(x)$ may have closed or informative expression, we list some copulas for example.

\begin{example}
	Assume $\rho(corr(Y,X_1),...,corr(Y,X_d))^T$ and $\Sigma_X$ denote the correlation matrix of X. If the copula of $(Y,X^T)^T$ is Gaussian Copula, then we have:
	$$m(x) = \mathbb{E}[F_0^{-1}(\Phi(u^T\Sigma_X^{-1}\rho+\sqrt{1-\rho^T\Sigma_X^{-1}\rho}Z))]$$
	where $u = (\Phi^{-1}(F_1(x_1),...,\Phi^{-1}(F_d(x_d))))^T$ and $Z\sim \mathcal{N}(0,1)$

\end{example}
\subsection{General Copula Regression method}\label{sec:subsamplingalg}
Linear models or Generalized linear model are popular statistical model for continous or categorical response prediction. As for short, we take linear regression model for short, generalized linear model can be viewed as a extension of linear regression model. Linear regression model often model the conditional expectation of $Y$ given $X$ as a linear function, however this may lead to lack of fit because of simplicity of liearn function. Another view on linear regression can be derived via generative modelling which leads us to consider copula regression and classification later. Assume variables $(X_1,...,X_d,Y)$ has a joint gaussian distribution with mean $\mu = (\mu_x^T,\mu_y)^T$ and covariance matrix $\Sigma = \begin{pmatrix}
\Sigma_{xx} & \Sigma_{xy}\\
\Sigma_{xy} & \Sigma_{yy}\\
\end{pmatrix}$,then $$\mathbb{E}(Y|X) = \mu_y - \Sigma_{yx}\Sigma_{xx}^{-1}(\mu_x - X)$$
So under assumption $(X_1,...,X_d,Y)$ has joint gaussian distribution, the prediction function is natrually a linear function. Evidences have been proposed that in many real world problem, joint distribution of covariates and response is far from gaussiann \cite{Rainer2012}. \cite{Embrechts2002} show how the Pearson correlation coefficient can be misleading when the underlying distributions are not normal. They advise using copulas to model data that are not normal because such models capture a greater variety of relationships (essentially being nonparametric). So there is need for statistic models to caputure more complex dependence structure among variables. \cite{Noh2013} have proposed a generic method to use copula dependence and thus fit regression model. In his work he propsed to fit model in a semi-parametric way. Other methods for fitting model have been proposed either. \cite{Chib2007} and \cite{Mara2013} introduced Bayesian and likelihood estimation methods based on penalized splines.e, \cite{Rainer2012} discussed a modification of the recursive bivariate probit that maintains the Gaussian assumption for the marginal distributions of the two equations while introducing non-Gaussian dependence between them using the Frank and
Clayton copulas. However, these metods only consider bivariate case while multivariate case is largely different from it. In this part, we do an empirical study of copula regression model in \cite{Noh2013} and discuss its pros and cons, while in latter section we will derive our copula regression model with binary outcome motivated by \cite{Noh2013}.

Assume $(X_1,...,X_d,Y)$ has copula smooth $C_\theta(u_1,...,u_d,v)$, and marginal cumulative distribution(density) $F_i(f_i)$. Noh2013 proposed to esitmiate marginal distribution with non-parametric method while fit maximum likelihood for parametric copula family. More clearly, they use kernel smoothed estimation $$\bar{F}_j(x_i) = \frac{1}{n}\Sigma \mathcal{K}(\frac{x_i-X_{i,j}}{h})$$ to be estimation of marginal distribution. As for copula estimation, there are a bunch of method which can estimate copula's parameter. Nonparametric methods for estimating c include kernel smoothing estimators (see for example \cite{Gijbels1990}, \cite{Charpentier2006} and \cite{Chen2010}) and Bernstein estimator (see \cite{Bouezmarni2013}). In spite of the great flexibility of nonparametric methods, they are typically affected by the curse of
dimensionality and they come with the difficult problem of selecting a good smoothing parameter. On the other hand imposing a parametric structure on both the copula and marginal distributions can lead to severely biased and inconsistent (fully parametric) estimator in case of misspecification. So a non-parametric marginal together with a parametric copula estimation is supposed to be considered.

\subsection{Simulation Study for Copula Regression}
The objective of this section is to compare the semi-parametric copula regression estimator proposed by Noh2013 with OLS both when the true copula family is known and when the copula family and its parameters are adaptively selected using the data. To this end, we consider the following data generating procedures (DGPs):
\begin{itemize}
	\item \textbf{DGP I.a} $(F_0(Y),F_1(X_1))\sim$ Clayton copula with parameter $\delta =1$; $Y \sim \mathcal{N}(\mu_Y=1,\sigma_Y^2=1)$, $X_1 \sim \mathcal{N}(\mu_{X_1}=0,\sigma_{X_1}^2=1)$. The resulting regression function is $m(x_1)=\mu_Y+\mathbb{E}[\sigma_Y\Phi^{-1}(T^{-1/\delta})]$, where $T \sim  f_T(t)=(1/\delta+1)(1+\xi)^{(1/\delta+1)}/(t+\xi)^{(1/\delta+2)}$ for $t>1$ and $\xi=F_{X_1}(x_1)^{-\delta}-1$.
	
	\item \textbf{DGP I.b}
	$(F_0(Y),F_1(X_1))\sim$ FGM copula with parameter $\theta=0.8$; $Y \sim  \mathcal{N}(\mu_Y=0,\sigma_Y^2=1)$, $X_1$ is generated from the Gumbel distribution $F_{X_1}(x_1)=1-exp(-exp(x_1))$. The resulting regression function is $m(x_1)=\mu_Y-\frac{\theta}{\sqrt{\pi}}\sigma_Y+2\frac{\theta}{\sqrt{\pi}}\sigma_YF_1(x_1)$.
	
	\item \textbf{DGP I.c}
	$(F_0(Y),F_1(X_1),...F_d(X_d))\sim$ Gaussian copula with correlation matrix $\Sigma =\bigl[ \begin{smallmatrix} 1 & \rho^T \\ \rho & \Sigma_X \end{smallmatrix} \bigr]$, where $\rho$ is a d-dimensional vector; $Y\sim \mathcal{U}(0,1)$; $X_j \sim \mathcal{N}(\mu_{X_j}=0,\sigma_{X_j}^2=1),j=1,...d$, $d=3$. We choose correlation matrix as $\Sigma =\bigl( \begin{smallmatrix} 1 & 0.23 & 0.90 & 0.67 \\ 0.23 & 1 & 0.51 & 0.26 \\ 0.90 & 0.51 & 1 & 0.49 \\ 0.67 & 0.26 & 0.49 & 1 \end{smallmatrix} \bigr)$. The resulting regression function is $m(x)=\Phi(\sum_{j=1}^d \frac{a_j}{\sqrt{2-\rho^Ta}}\Phi^{-1}(F_j(x_j)))$, where $a=(a_1,...,a_d)^T\equiv{\Sigma_X^{-1}\rho}$.
	
	\item \textbf{DGP II.a} the same as DGP I.c
	\item \textbf{DGP II.b} $(F_0(Y),F_1(X_1),...F_d(X_d))\sim$ R-Vine copula with the same structure and parameters as the illustrating example in the help page of function RVineMatrix of R package \emph{VineCopula}; $Y\sim \mathcal{U}(0,1)$; $X_j \sim \mathcal{N}(\mu_{X_j}=0,\sigma_{X_j}^2=1),j=1,...d$, $d=4$.
	\item \textbf{DGP II.c} $(F_0(Y),F_1(X_1),...F_d(X_d))\sim$ Clayton copula with parameter $\delta=1$; $Y$ is generated from the Beta distribution with parameters $\alpha=0.5,\beta=0.5$; $X_j \sim \mathcal{N}(\mu_{X_j}=0,\sigma_{X_j}^2=1),j=1,...d$, $d=2$.
	\item \textbf{DGP II.d} $(F_0(Y),F_1(X_1),...F_d(X_d))\sim$ T copula with correlation matrix $\Sigma$ as DGP I.c and degree of freedom $df=5$; $Y$ is generated from the Beta distribution with parameters $\alpha=0.5,\beta=0.5$; $X_j \sim \mathcal{N}(\mu_{X_j}=0,\sigma_{X_j}^2=1),j=1,...d$, $d=3$.
	
\end{itemize} 
As mentioned above, we conduct two set of simulation studies. In the first part, data are generated from DGP I.a to DGP I.c, and when we estimate copula regression, we know the true copula structure and estimate copula parameters using pseudo-MLE. In the second part, data are generated from DGP II.a to DGP II.d, and when we estimate copula regression, we adaptively select copula structure and parameters from data. In all seven experiments, we do simulations $N=200$ times, each time with a data sample of $n=100$ observations. Then we calculate IMSE, IBIAS and IVAR in a fixed evaluation set with $I=150$ observations in each experiment as follows:
$$IMSE=\frac{1}{N}\sum_{l=1}^N ISE(\hat{m}^{(l)})\equiv{\frac{1}{N}\sum_{l=1}^N[\frac{1}{I}\sum_{i=1}^I (\hat{m}^{(l)}(x_i)-m(x_i))^2]}$$
$$=\frac{1}{I}\sum_{i=1}^I (m(x_i)-\bar{\hat{m}}(x_i))^2 + \frac{1}{I}\sum_{i=1}^I[\frac{1}{N}\sum_{l=1}^N (\hat{m}^{(l)}(x_i)-\bar{\hat{m}}(x_i))^2] \equiv{IBIAS+IVAR}$$
where $\{(y_i,x_i),i=1,...I \}$ is a fixed evaluation set, which corresponds to a random sample of size $I=150$ generated from the DGP, $\hat{m}^{(l)}(\cdot)$ is the estimated regression function from the $l$th data sample and $\bar{\hat{m}}(x_i)=N^{-1}\sum_{l=1}^N\hat{m}^{(l)}(x_i)$. We should point out that in the second part of simulation study, multi-dimensional $X$ makes it difficult to calculate the true regression function $m(x)$. So we replace $m(x_i)$ with $y_i$ in the calculation of IMSE and IBIAS. Then the variances of error terms $y_i-m(x_i)$ are included in IMSE and IBIAS, making them larger than their counterparts in the first part of simulation study.

\subsubsection{Simulation Study I: Known Copula Structure}
In this part, data are generated from DGP I.a to DGP I.c and when we estimate copula regression, we know the true copula structure and then estimate parameters using pseudo-MLE. Table 1 shows the IMSE together with the IBIAS and the IVAR of copula regression and OLS (with intercept). In all three settings, copula regression has much lower bias but higher variance than OLS. Together, copula regression attains lower IMSE. This simulation study reveals the potential of copula regression. But to apply it practically, we need to adaptively select copula structure and parameters from data, which is dealt with in the following section.

\begin{table}[htbp]
	\centering
	\caption{Copula regression and OLS: Known Copula Structure}
	\begin{tabular}{cccccccc}
		\hline
		\multirow{2}{*}{Y margin} & \multirow{2}{*}{copula}
		&\multicolumn{2}{c}{IMSE} & \multicolumn{2}{c}{IBIAS} & \multicolumn{2}{c}{IVAR} \\ 
		
		& & copula & OLS & copula & OLS & copula & OLS \\
		\hline
		\multirow{2}{*}{normal} & Clayton & 0.0197 & 0.0611 & 0.0031 & 0.0463 & 0.0166 & 0.0147 \\
		& FGM & 0.0159 & 0.0241 & 0.0006 & 0.0068 & 0.0153 & 0.0173 \\
		\hline
		Uniform & Gaussian & 0.0010 & 0.0037 & 0.0001 & 0.0033 & 0.0009 & 0.0004 \\
		\hline
	\end{tabular}
\end{table}

\subsubsection{Simulation Study II: Unknown Copula Structure}
In this part, data are generated from DGP II.a to DGP II.d and we adaptively select copula structure and parameters from data. This step may be difficult especially when the number of covariates is large. The reason is that the set of high-dimensional copulas available in the literature is limited to very special and restrictive copula families such as elliptical copulas and Archimedean copulas. For this reason, we make use of the recent available work about the simplified pair-copula decomposition. The main idea is to decompose a multivariate copula to a cascade of bivariate copulas so that we can take advantage of the relative simplicity of bivariate copula selection and estimation. In our simulation, we choose one decomposition (R-Vine structure) for the data and use R package \emph{VineCopula} to select copula structure and estimate parameters. \\
Again, we compare copula regression with OLS (with intercept). In all four settings, copula regression has lower bias but higher variance than OLS. Together, copula regression attains lower IMSE. Note that now the variances of error terms are also included in IMSE and IBIAS, so they are higher in the Gaussian setting compared to I.c. This simulation study illustrates that copula regression may have higher prediction power than OLS practically and in the next section, we will show some evidence in the real data.

\begin{table}[htbp]
	\centering
	\caption{Copula regression and OLS: Unknown Copula Structure}
	\begin{tabular}{cccccccc}
		\hline
		\multirow{2}{*}{Y margin} &\multirow{2}{*}{copula} &\multicolumn{2}{c}{IMSE} & \multicolumn{2}{c}{IBIAS} & \multicolumn{2}{c}{IVAR} \\ 
		
		& & copula & OLS & copula & OLS & copula & OLS \\
		\hline
		\multirow{2}{*}{Uniform} &
		Gaussian & 0.0077 & 0.0078 & 0.0054 & 0.0074 & 0.0023 & 0.0004 \\
		
		& R-Vine & 0.0116 & 0.0136 & 0.0083 & 0.0129 & 0.0033 & 0.0007 \\
		\hline
		\multirow{2}{*}{Beta(0.5,0.5)} &
		Clayton & 0.0887 & 0.0910 & 0.0854 & 0.0881 & 0.0033 & 0.0029 \\
		
		& T & 0.0126 & 0.0204 & 0.0082 & 0.0191 & 0.0043 & 0.0013 \\
		\hline
	\end{tabular}
\end{table}

\subsection{Real Data Study for Copula Regression}
In this section, we analyze Boston Housing Data. The data consist of 506 observations with 14 variables. The dependent variable is MEDV, the median value of owner-occupied homes in \$1000's. The independent variables include per capita crime rate, nitric oxides concentration, weighted distances to five Boston employment centres, average number of rooms per dwelling, index of accessibility to radial highways, property-tax rate, pupil-teacher ratio etc. To estimate regression function, we consider 4 methods:
\begin{itemize}
	\item (OLS) Least-squre estimator
	\item (GAM) Generalized additive model estimator
	\item (CART) Classification and Regression Tree
	\item (CR) Copula regression method
\end{itemize}

We use a smoothing spline to fit GAM using function \emph{gam} of R package \emph{mgcv}.For CART, we use R package \emph{rpart} and choose complexity parameter equals 0.001. As an evaluation measure of each estimator, we randomly split the data into training set ($n=337$) and testing set ($n=169$) for 100 times and calculate the mean and standard error of MSE in the test set for each estimator. Table 3 shows that, copula regression can attain much lower prediction error than OLS, slightly lower than CART and almost the same as GAM. As for the stability of prediction power, copula regression has slightly higher MSE standard error than OLS and CART, but much lower than that of GAM. All together, copula regression can balance prediction precision and stability and do fairly good job in the real data.

\begin{table}[htbp]
	\centering
	\caption{Real Data Comparison of OLS, GAM, CART and CR}
	\begin{tabular}{ccccc}
		\hline
		
		MSE & OLS & GAM & CART & CR \\
		\hline
		mean & 0.465 & 0.295 & 0.318 & 0.296\\
		\hline
		sd & 0.0670 & 0.2302 & 0.0711 & 0.0781 \\
		\hline
	\end{tabular}
\end{table}

\section{Binary outcome model}\label{sec:appr-optim-subs}
\subsection{Latent Variable formulation}
In this section, we give our formulation of Binary outcome regression model along with our proposed sampling based fitting method. Recape last section, we give a copula regression model and corresponding semi-parametric fitting method. Evidence have shown the model can perform well if both covariate and response are continous. In real world applications, variables with binary outcome play an important role. In medical field, doctor uses patients' observed varaible to predict whether a patient has certain disease. In individual credit risk management field, bank uses custormers' observed variable to judge if a custormer may default in near future or not. There are cases revealed the importance of prediction for binary response. Classical model such as logistic regression or linear discriminant analysis has been proposed for prediction of binary variables. However the simplicity of linear fomula reduces the dependency structure for varaibles and will lead to lack of fit. In this paper, we consider use a latent variable model with copula to model varaible dependency. Assume we have $(X_1,...,X_n,Y)$ are observed data where $Y$ takes value from $\{0,1\}$ and $X_i$ are continous variable taking values in $\mathcal{R}$. Assume the relation ship between $X$ and $Y$ are connected from one latent variable $Z\in [0,1]$, for which:
$$Y|X,Z \equiv Y|Z \sim Ber(Z)$$
$$(X,Z)\sim c_\theta(F_X(x),F_Z(z))f_X(x)f_Z(z)$$
where $c_\theta(u,v)$ is the \textbf{Copula Density} of $(X,Z)$. The nature of the model can be interpreted as, the response $Y$ is determind by latent probability $Z$ via a bernoulli experiment $Y|X \sim Ber(Z)$, while the covariates $X$ and latent probability $Z$ share joint distribution density $c_\theta(F_X(x),F_Z(z))f_X(x)f_Z(z)$. Our attempt to use latent variable to reveal the connection between covariates and binary outcome is not the first one. \cite{Rainer2012} and \cite{Radice2016} have proposed a latent probit model with copula dependency, but they assumed a gaussian latent variable $Z$ which does not have much explainable meaning. To our best knowledge, we are the first one to use a bernoulli response with latent probability variable with copula to represent dependency. It is benefitial to use bernoulli response to model the binary outcomes. One benefit is if we assume the marginal distribution of latent probability has a beta distribution form, the nature property of \textbf{one peak} for beta distribution can interpret the prior response strength, while covaraites $X$ are then to adjust the response strength. The second benifit is once we fit the parameter of the model, the natural conditonal mean can $\mathbb{E}(Z|X)$ is the prediction probability for an individual obeservation. The function $m(x) = \mathbb{E}(Z|X=x)$ can not only predict response but also the probability with which the response will take $1$ or $0$. The probability is of great importance in many statistical application such as individual credit scoring or custormer click rate prediction. In the following part we will give our proposed methods for fitting the model.

\subsection{Fitting Alogorithm}
Assume variables $(X,Y),X\in\mathcal{R},Y\in\{0,1\}$ are observed variable, $Z\in [0,1]$ is the latent variable. Assume $Y|Z,X \sim Ber(Z)$ and $(X,Z)\sim c_\theta(F_X(x),F_\phi(z))f_X(x)f_\phi(z)$. The joint density of $(X,Y,Z)$ is 
$$\mathbb{P}(x,y,z) = p(x,z)p(y|z) = c_\theta(F_X(x),F_\phi(z))f_X(x)f_\phi(z)\times z^y(1-z)^{1-y}$$
The likelihood for parameter $(\theta,\phi)$ is
$$L(\theta,\phi) = p(x,y) = \int c_\theta(F_X(x),F_\phi(z))f_X(x)f_\phi(z)\times z^y(1-z)^{1-y} dz$$

The derivative for likelihood wrt parameters are under regularity condition:
$$\frac{\partial L(\theta,\phi)}{\partial \theta} = \int\frac{\partial c_\theta(F_X(x),F_\phi(z))}{\partial \theta}f_X(x)f_\phi(z)\times z^y(1-z)^{1-y} dz$$

$$\frac{\partial L(\theta,\phi)}{\partial \phi} = \int[\frac{\partial c_\theta(u,v)}{\partial v}\frac{\partial F_\phi(z)}{\partial \phi}|_{v=F_\phi(z)}^{u=F_X(x)} + c_\theta(F_X(x),F_\phi(z))\frac{\frac{\partial f_\phi(z)}{\partial \phi}}{f_\phi(z)}] \times f_X(x) z^y(1-z)^{1-y}f_\phi(z)dz$$

For $F_X(.)$, we can use non-parametric method like kernel smoothing method to estimate. It is clear in most case the integral will not have explicit formula, but one fortunate thing is that because of good structure of model, the integral can be interpreted as a expectation for $Z$ if $Z$ have density $f_\phi(z)$. The fact means under current parameters $(\theta,\phi)$, one can sample $(z_1,...,z_K)\sim f_\phi(z)$ and use sample mean to estimate true parameters. 
$$\frac{\hat{L}(\theta,\phi)}{\theta} = [\Sigma_{k=1}^K \frac{\partial c_\theta(F_X(x),F_\phi(z_k))}{\partial \theta}f_X(x)\times z_k^y(1-z_k)^{1-y}]/K$$

$$\frac{\hat{L}(\theta,\phi)}{\phi} = [\Sigma_{k=1}^K\frac{\partial c_\theta(u,v)}{\partial v}\frac{\partial F_\phi(z_k)}{\partial \phi}|_{v=F_\phi(z_k)}^{u=F_X(x)} + c_\theta(F_X(x),F_\phi(z_k))\frac{\frac{\partial f_\phi(z_k)}{\partial \phi}}{f_\phi(z_k)}] \times f_X(x) z_k^y(1-z_k)^{1-y}/K$$

In practice, the likelihood is strictly bounded in $[0,1]$, thus the derivative may vanish because of computation accuracy and prevent the algorithms to converge. Tt is a usual practice to estimate gradients of log likelihood, and we briefly give the formula here:
$$\frac{\partial l(\theta,\phi)}{\partial \theta} = \frac{\int\frac{\partial c_\theta(F_X(x),F_\phi(z))}{\partial \theta}f_\phi(z)\times z^y(1-z)^{1-y} dz}{\int c_\theta(F_X(x),F_\phi(z))f_\phi(z)\times z^y(1-z)^{1-y} dz}$$

$$\frac{\partial l(\theta,\phi)}{\partial \phi} = \frac{\int[\frac{\partial c_\theta(u,v)}{\partial v}\frac{\partial F_\phi(z)}{\partial \phi}|_{v=F_\phi(z)}^{u=F_X(x)} + c_\theta(F_X(x),F_\phi(z))\frac{\frac{\partial f_\phi(z)}{\partial \phi}}{f_\phi(z)}] \times z^y(1-z)^{1-y}f_\phi(z)dz}{\int c_\theta(F_X(x),F_\phi(z))f_\phi(z)\times z^y(1-z)^{1-y} dz}$$

where $l(\theta,\phi) = \log [\int c_\theta(F_X(x),F_\phi(z))f_\phi(z)\times f_X(x)z^y(1-z)^{1-y} dz]$

Our algorithm is show as:\\
\begin{algorithm}[H]
	\caption{Sampling Gradient Fitting Algorithm}
	\KwIn{Observed Data $(X_i,Y_i),1\leq i\leq N$}
	\KwOut{Fitted Parameter $(\hat{\theta},\hat{\phi})$,stepsize $\epsilon$}
	
	initialize $(\theta^0,\phi^0)$\\
	
	\For{t in $1:Max\_Iter$}{
		Sample $N\times K$ samples $(z_{nk})$\\

		$\hat{g_\theta} = \frac{\Sigma_{nk}\frac{\partial c_\theta(F_X(x_n),F_\phi(z_{nk}))}{\partial \theta}\times z_{nk}^{y_n}(1-z_{nk})^{1-y_n}}{\Sigma_{nk} c_\theta(F_X(x_n),F_\phi(z_{nk}))\times z_{nk}^{y_n}(1-z_{nk})^{1-y_n} }$\\
		
		$\hat{g_\phi} = \frac{\Sigma_{nk}[\frac{\partial c_\theta(u,v)}{\partial v}\frac{\partial F_\phi(z)}{\partial \phi}|_{v=F_\phi(z_{nk})}^{u=F_X(x_n)} + c_\theta(F_X(x_n),F_\phi(z_{nk}))\frac{\frac{\partial f_\phi(z_{nk})}{\partial \phi}}{f_\phi(z_{nk})}] \times z_{nk}^{y_n}(1-z_{nk})^{1-y_n}}{\Sigma_{nk} c_\theta(F_X(x_n),F_\phi(z_{nk}))\times z_{nk}^{y_n}(1-z_{nk})^{1-y_n}}$\\
		
		$\theta^{t+1} = \theta^{t} + \epsilon \hat{g_\theta}$\\
		
		$\phi^{t+1} = \phi^{t} + \epsilon \hat{g_\phi}$\\
		
	}
\end{algorithm}

\subsection{Some examples of the method}\label{sec:mV}
In this section, we give some examples with different $c_\theta(u,v)$ and $f_\phi(z)$ to get explicit update formula with our algorithm.

\textbf{Proposition I} When $c_\theta$ is Guassian copula, we have 
$$\frac{\partial l(\Sigma,\phi)}{\partial \Sigma}=\frac{\int z^y (1-z)^{1-y}c_\Sigma(F_X(x),F_\phi(z))[\frac{1}{2}\Sigma^{-1}tt^T\Sigma^{-1}-\frac{1}{2}\Sigma^{-1}]f_\phi(z)dz}{\int z^y (1-z)^{1-y}c_\Sigma(F_X(x),F_\phi(z))f_\phi(z)dz}$$ 
where $t=(\Phi^{-1}(F_\phi(z)),\Phi^{-1}(F_{X_1}(x_1)),...\Phi^{-1}(F_{X_d}(x_d)))^T$, $\Sigma$ is the correlation matrix of Guassian copula.


\subsection{Simulation Studies}\label{sec:Poisson-sim}
This section aims to evaluate the proposed binary output copula regression method. We compare this new method with logit regression to gain some insight about its performance. To this end, we consider the following data generating procedures (DGPs):
\begin{itemize}
	\item \textbf{DGP III.a} $(F_0(Z),F_1(X_1),...F_d(X_d))\sim$ Clayton copula with parameter $\delta =1$; $Z \sim \mathcal{U}(0,1)$, $X_j \sim \mathcal{N}(\mu_{X_j}=0,\sigma_{X_j}^2=1)$, $d=3$. The outcome $Y$ is subject to Bernoulli distribution with success rate $Z$.
	\item \textbf{DGP III.b}
	The structure of $(Z,X_1,...X_d)$ is the same as DGP I.c. The outcome $Y$ is subject to Bernoulli distribution with success rate $Z$.
	\item \textbf{DGP III.c} $(X_1,...X_d),d=4$ is generated from multivariate normal distribution which has correlation matrix as in DGP I.c and standard normal marginal distribution. Then we generate $Z=sigmoid(X\beta)$, where $\beta=(1,-1,-1,1)^T$. The outcome $Y$ is subject to Bernoulli distribution with success rate $Z$.
\end{itemize}
We first generate a data sample of $n=300$ observations from DGP III.a to DGP III.c. After that, we randomly split the data sample into training set (200 obs) and testing set (100 obs). We estimate binary-outcome copula regression and logit regression on the training set and then calculate AUC and KS-value for them on the testing set. Table 4 below shows the average AUC and KS-value for the two method. On average, our binary-outcome copula regression method attains slightly better AUC and KS-value than logit regression.

\begin{table}[htbp]
	\centering
	\caption{BOCR and Logit}
	\begin{tabular}{ccccc}
		\hline
		&\multicolumn{2}{c}{AUC} & \multicolumn{2}{c}{KS value}\\
		& logit & BOCR & logit & BOCR \\
		\hline
		Clayton & 0.639 & 0.651 & 0.284 & 0.295\\
		\hline
		Guassian & 0.747 & 0.742 & 0.375 & 0.408 \\
		\hline
		Logit & 0.604 & 0.612 & 0.218 &0.242 \\
		\hline
	\end{tabular}
\end{table}

\subsection{Real Data Studies} \label{sec:realdata}
In this section, we analyze Breast-Cancer-Wisconsin Data. The data consist of 699 observations with 10 variables. The outcome variable is CLASS, whether the cancer is benign or malignant. The independent variables include clump thickness, uniformity of cell size and shape, marginal adhesion etc. We consider 4 classification algorithms: 
\begin{itemize}
	\item (Logit) Logit regression 
	\item (CART) Classification and Regression Tree
	\item (SVM) Supporting Vector Machine
	\item (BOCR) Binary-Outcome Copula Regression
\end{itemize}

We use R package \emph{rpart} to estimate CART and package \emph{e1071} to estimate SVM. We randomly split the data into training set (n=466) and testing set (n=233) and calculate AUC and KS-value in the testing set for each method. Table 5 shows that, BOCR can attain similar AUC and KS-value as the other three methods. From the results, it seems that it is fairly easy to classify cancer to benign or malignant since all four methods attain KS-value higher than 0.9! 

\begin{table}[htbp]
	\centering
	\caption{Real Data Analysis for BOCR}
	\begin{tabular}{ccccc}
		\hline
		
		& logit & CART & SVM & BOCR \\
		\hline
		AUC & 0.9956 & 0.9780 & 0.9964 & 0.9964\\
		\hline
		KS value & 0.9482  & 0.9350 & 0.9539 & 0.9548 \\
		\hline
	\end{tabular}
\end{table}


\section{Conclusion and Future Work}
In this paper, we empirically investigated copula regression model and proposed a binary outcome copula regression model. We give a sampling gradient method for fitting the model together with simulation study and real data study of the model. Evidence has show our model can overwhelm logistic regression model and machine learning models such as CART and SVM. To our best knowledge, our model is the first copula based model to deal with multivariate variables with single binary outcome. Also we are the first attempt to introduce Sampling Gradient Estimation in fitting such models. However, there are still manys work to do in future. We briefly discuss two aspects.

One future work may be evaluation of various copula functions under our model framework. In our paper we mainly focus on Gaussian Copula as a template, which means nearly all other copulas can be evaluated in the same way we do.

Another future work may be Variance Reduction in Sample Gradient Estimation Procedure. One knows their are bunches of methods to reduce variance for Monte Carlo Estimation, some of them include importance sampling, Rao-Blackwellization et-al. The application of variance reduction tricks in our model will be an interesting research direction.

\section*{Supplementary Material}

\par
\section*{Acknowledgements}
This paper is motivated by our final project for \textbf{Multivariate Statistics} set up by Guanghua School of Management, Peking University, 2020 fall. In class, professor Chen Songxi have offered great help for us. In this section, we want to say sincere thanks to professor Chen on both his profound knowledge and his enthusiastic help.

\par


\bibhang=1.7pc
\bibsep=2pt
\fontsize{9}{14pt plus.8pt minus .6pt}\selectfont
\renewcommand\bibname{\large \bf References}
\expandafter\ifx\csname
natexlab\endcsname\relax\def\natexlab#1{#1}\fi
\expandafter\ifx\csname url\endcsname\relax
  \def\url#1{\texttt{#1}}\fi
\expandafter\ifx\csname urlprefix\endcsname\relax\def\urlprefix{URL}\fi


\newpage
\vskip .1cm
\noindent
School of Mathematical Sciences, Peking University, Beijing 100871, China.
\vskip 2pt
\noindent
E-mails: luoweijian@math.pku.stu.edu.cn
\vskip 2pt

\noindent
School of International Studies, Peking University, Beijing 100871, China.
\vskip 2pt
\noindent
E-mail: maiwo@nsd.pku.edu.cn
\vskip 2pt

\end{document}